\documentclass[twocolumn,twocolappendix]{aastex631}

\usepackage{latexsym}
\usepackage{graphicx}
\usepackage{amsmath}           % AMS extended math package
\usepackage{graphicx}          % EPS inclusion
\usepackage{txfonts}           % fonts
\usepackage{mathrsfs}          % Raph's Smith Formal Script
\usepackage{rotating}          % rotates pages 
\usepackage{subfigure}         % multiple figures
\usepackage{afterpage}         % improves control float objects
\usepackage{xspace}            % gentle spacing after macro
\usepackage{natbib}            % package for c itations
\usepackage{hyperref}          % link reference
\usepackage{amssymb}
\usepackage{chemformula}       
\usepackage{longtable}
\usepackage{epsf}
\let\ce\ch
\usepackage{ulem}
\usepackage{xcolor}
\usepackage{textcomp}
\DeclareUnicodeCharacter{2212}{\textendash}

\begin{document}

\title{A high-resolution spectroscopic analysis of aminoacrylonitrile and an interstellar search towards G+0.693}

\author[0000-0002-0786-7307]{D.~Alberton}
\affiliation{Center for Astrochemical Studies, Max Planck Institute for Extraterrestrial Physics, Gie\ss enbachstr.~1, 85748 Garching, Germany}

\author{V. Lattanzi}
\affiliation{Center for Astrochemical Studies, Max Planck Institute for Extraterrestrial Physics, Gie\ss enbachstr.~1, 85748 Garching, Germany}

\author{C. Endres}
\affiliation{Center for Astrochemical Studies, Max Planck Institute for Extraterrestrial Physics, Gie\ss enbachstr.~1, 85748 Garching, Germany}

\author{V. M. Rivilla}
\affiliation{Centro de Astrobiolog\'ia (CSIC-INTA), Ctra. de Ajalvir Km. 4, Torrej\'on de Ardoz, 28850 Madrid, Spain}

\author{J.C. Guillemin}
\affiliation{Univ Rennes, Ecole Nationale Supérieure de Chimie de Rennes, CNRS, ISCR – UMR6226, F-35000 Rennes, France}

\author{P. Caselli}
\affiliation{Center for Astrochemical Studies, Max Planck Institute for Extraterrestrial Physics, Gie\ss enbachstr.~1, 85748 Garching, Germany}

\author{I. Jim\'enez-Serra}
\affiliation{Centro de Astrobiolog\'ia (CSIC-INTA), Ctra. de Ajalvir Km. 4, Torrej\'on de Ardoz, 28850 Madrid, Spain}

\author{J. Mart\'in-Pintado}
\affiliation{Centro de Astrobiolog\'ia (CSIC-INTA), Ctra. de Ajalvir Km. 4, Torrej\'on de Ardoz, 28850 Madrid, Spain}


\begin{abstract}
Cyanides, ranging from three carbon atoms to PAHs, and alkenyl compounds are abundant in the interstellar medium (ISM). Aminoacrylonitrile (3-Amino-2-propenenitrile, \ch{H2N-CH=CH-C+N}), an alkenyl cyanide, thus represents a promising candidate for new interstellar detection.
A comprehensive spectroscopic laboratory investigation of aminoacrylonitrile in its rotational ground vibrational state has been herein performed. The measurements carried out up to the THz regime made it possible to generate a precise set of reliable rest frequencies for its search in space up to sub-millimetre wavelengths.
The \textit{Z}-aminoacrylonitrile ($Z$-apn) isomer spectrum has been recorded employing a source-modulated sub-millimetre spectrometer, from 80 GHz to 1 THz. A combination of Doppler and sub-Doppler measurement regimes allowed to record 600 new lines.
The collected data have enabled the characterisation of a set of spectroscopic parameters up to decic centrifugal distortion constants. The catalogue generated from the improved spectral data has been used for the search of $Z$-apn in the spectral survey of the G+0.693-0.027 molecular cloud located in the central molecular zone, in the proximity of the Galactic centre.
\end{abstract}

	
%########################################################################################

\section{Introduction} \label{sec:intro}

Scientists have proposed many theories in order to explain the origin of life. According to one of the most widely accepted, it may have originated from the self-catalytic function of an RNA chain, an idea originally proposed by \citealt{gilbert1986} and currently known as the RNA-world. However, how the RNA and subsequently life were formed still remains unknown. Among the various hypotheses, it is possible to outline two main lines of thought. Life could have emerged starting from precursors of biomolecular building blocks synthesised \textit{in situ}, during our planet's cooling. Successive RNA polymerization could have then emerged in warm little ponds as a result of condensation of lightning synthesised abiotic precursors \citep{pearce_2022rna}, produced in a Urey-Miller experiment fashion \citep{Stanley_1953}.
Another hypothesis is that prebiotic compounds might have had an extraterrestrial origin. The prebiotic material could have been formed in the interstellar medium (ISM), and subsequently delivered to Earth to give rise to more complex molecules (\citealt{Oparin_1957}, \citealt{Chyba_Sagan_1992}). Pristine building blocks synthesized in the ISM would have been inherited by interplanetary dust particles, comets, and meteorites that in later stages might have acted as carriers of those seeds essential for life ignition (\citealt{Chyba_1990}, \citealt{altwegg2016}, \citealt{rivilla2020}). 
Only in the last stages, molecules might have started to self-organize and replicate on Earth, giving rise to life as we know it.

\noindent
Clues about prebiotic precursors' origin come from the composition of celestial bodies. Carbonaceous chondrites, considered among the youngest objects formed along with the formation of planetary systems, have been found rich in sugar-like compounds, long-chain monocarboxylic acids, amino acids, and nucleobases precursors, such as purines and pyrimidines (\citealt{Stoks_1979}, \citealt{Pizzarello_2006}, \citealt{Burton_2012}, \citealt{Oba_2022}). Following the work by \citet{Bockelee-Morvan_2000}, further evidence of the partial inheritance of volatile composition from prestellar and protostellar evolution phases to cometesimals and planetesimals, came from the PILS\footnote{Protostellar Interferometric Line Survey.} project. With the data therein obtained, \citet{Drozdovskaya_2019} were able to outline the correlation existing among CHO-, N-, and S-bearing molecules between the low-mass protostar IRAS 16293-2422, a homologous of our protosolar system, and the bulk composition of the comet 67P/Churyumov-Gerasimenko. Analogously, \cite{oberg_2015} observed a correlation between abundances of interstellar Complex Organic Molecules (iCOMs) in comets and protoplanetary disks and many detections of several key precursors of the RNA world sustain such an idea (e.g \citealt{Zeng_2019}, \citealt{Jimenez-Serra_2020}, \citealt{rivilla2022a}).

Although the chemical processes of such systems are being studied extensively, our knowledge is still limited by what has been therein identified. Predictions of current abundances rely on observations, and since physical conditions are time-dependent we might still miss a large portion of information \citep{Herbst-vanDishoeck_2009}. The identification of new molecules can expand our chemical and physical insight of star-forming regions, and iCOMs have been proved to be excellent probes. Precise identification of ISM chemical species thus remains a crucial point to understand these environments.

\begin{figure}[tb]
 \centering
 \includegraphics[width=0.49\textwidth]{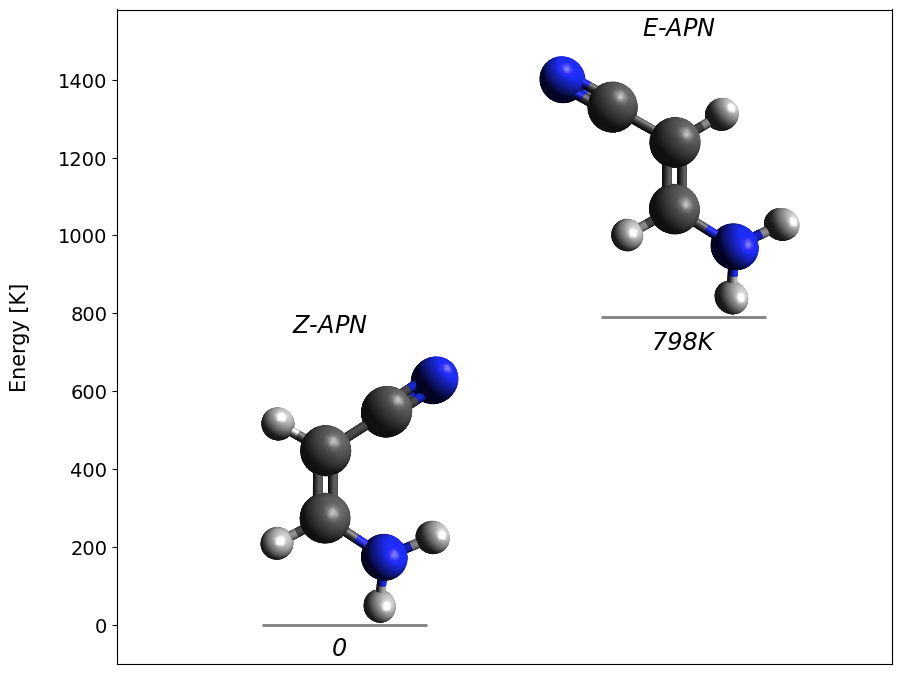}
 \caption{Energy level diagram of the two most stable isomers of 3-Amino-2-propenenitrile, $Z$-apn (left) and $E$-apn (right).}
    \label{fig:isom_energy}
\end{figure}

In this study, we focused on 3-Amino-2-propenenitrile (\ch{H2N-CH=CH-C+N}, Aminoacrylonitrile or apn, from here on), a promising candidate that could take part in the chemical network leading to amino-acids formation. Previous works \citep{Miller_1974,Sanchez_1966,Kitadai_2018} suggested that in the liquid phase, this molecule might be one of the key steps leading to two $\alpha$-amino acids Asparagine and Aspartic acid. Apn is the adduct of ammonia to cyanoacetylene (\ch{H-C+C-CN}, firstly detected in space by \citealt{Turner_1971}), and is a member of the ISM well represented cyanide group. 
Cyanides represent one of the most abundant families in the ISM, and are a key molecular precursors of prebiotic chemistry \citep{Rivilla_CN_2022}. In recent years several prebiotic species have been detected (e.g. cyanomethanimines by \citealt{zaleski2013detection}, \citealt{rivilla2019abundant} and glyconitrile by \citealt{Zeng_2019}).
In 2021 alone, among closed shell molecules, 1-cyano-1,3-cyclopentadiene (\ce{c-C5H5CN}, \citealt{McCarthy_2021}), \ce{HC11N} (\citealt{Loomis_2021}),  trans-cyanovinylacetylene and vinylcyanoacetylene (\ce{HC+CCH=CHC+N} and \ce{H2C=CHC3N}, respectively, \citealt{Lee_2021a}), 2-cyanocyclopentadiene (\ce{C5H5CN}, \citealt{Lee2021b}),  cyano thioformaldehyde (\ce{HCSCN}, \citealt{Cernicharo_2021}) and cyanoacetyleneallene (\ce{H2CCCHC3N}, \citealt{Shingledecker_2021}) were detected; among open shell molecules, cyanomidyl radical (\ce{HNCN}, \citealt{Rivilla_2021}), 3-cyano propargyl radical (\ce{CH2CCCN}, \citealt{Cabezas_2021}) and Magnesium radicals (\ce{MgC5N}, \citealt{Pardo_2021}) were detected.

\noindent
The importance of apn comes also from its prochiral trait, i.e. it can be converted into new and different chiral molecules in a single chemical step.  To date, only one chiral molecule has been detected in the ISM: propylene oxide (\ce{CH3CHCH2O}, \citealt{McGuire_2016}). Understanding the abundance of chiral molecules in space would make it possible also to explain the origin of enantiomeric abundance in meteorites \citep{Pizzarello_2011, de_Marcellus_2011} and hopefully shed light on the predominance of left-handed amino acids in living organisms.

The quiescent G+0.693-0.027 molecular clouds (hereafter G+0.693), located at the central molecular zone (CMZ) in the inner $\sim$500pc of our Galaxy, has recently shown to be abundant in complex organic molecules of prebiotic interest. Since 2019, it has been the subject of 13 new interstellar detections, including many cyanides \citep{Rivilla_CN_2022} and amines like NH$_2$OH \citep{rivilla2020b}, ethanolamine \citep{rivilla2021a}, vinyl and ethyl amine \citep{zeng2021}.
In a previous work carried out by \citet{Askeland_2006}, a low-frequency experiment characterised \textit{Z}-apn isomer in the 4-80 GHz band. However, the excitation temperature ($T_{\rm ex}$) found for most of the molecules in G+0.693 molecular cloud, is typically in the range 5-20 K (see e.g. \citealt{zeng2018}). At a temperature of 15 K, the \textit{Z}-apn Planck function picks at almost 75 GHz, where one of the most intense transitions, the $34_{6-28}-33_{6-27}$, has an uncertainty of almost 19 km s\textsuperscript{-1}, insufficient to guide an accurate search for this molecule in space. 
In the present work, we extended the spectroscopic characterisation of apn up to 1 THz. This frequency coverage and the accuracy of high resolution (43 kHz) spectroscopy measurements, allowed us to improve the precision of rotational and distortion constants producing a reliable catalogue for the search of this molecule not only in cold molecular clouds but in hotter ISM regions as well. It will therefore be possible to understand in which stages \textit{Z}-apn might have formed and how chemical complexity can differ among the different phases of the star formation.

%########################################################################################
\begin{figure}[tb!]
 \centering
 \includegraphics[width=0.5\textwidth]{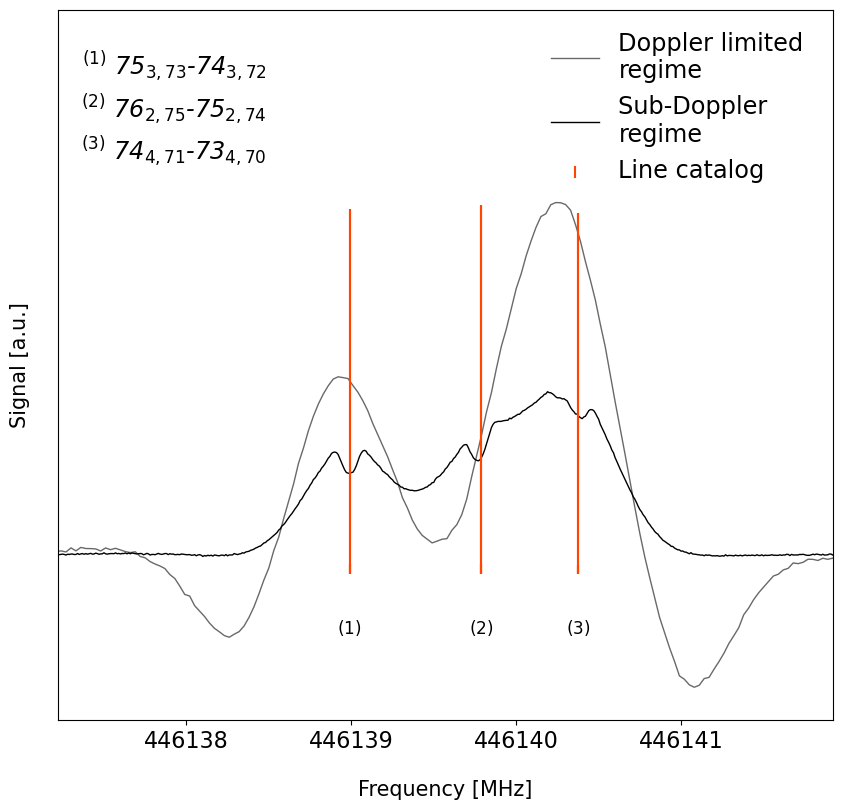}
 \caption{Doppler and Sub-Doppler measurements comparison for $a$-type \textit{Z}-apn isomer transitions $J'$$_{K_a',K_c'}$-$J_{K_a,K_c}$ = $75_{3-73}$-$74_{3-72}$, $76_{2-75}$-$75_{2-74}$ and $74_{4-71}$-$73_{4-70}$. For the Sub-Doppler regime, modulation width: 100 kHz; modulation frequency: 15 kHz; total integration time: 555 s. For Doppler-limited conditions, modulation width: 600 kHz; modulation frequency: 40kHz total; total integration time: 88s.}
    \label{fig:lamb_dip}
\end{figure}

\section{Experiment}\label{sec:exp}

\noindent
The spectroscopic characterization of $Z$-apn, was conducted using the CASAC (Center for Astrochemical Studies Absorption Cell) spectrometer of the Center for Astrochemical Studies at the Max Planck Institute for Extraterrestrial Physics. 
The instrument has been described comprehensively elsewhere \citep{Bizzocchi2017AccurateSR}, and only the distinctive features of the instrument are listed below. 

The main radiation source of the CASAC spectrometer is a frequency synthesizer (Keysight E8257D) locked to a Rb atomic clock (Stanford Research Systems) operating at 10 MHz for accurate frequency and phase stabilisation. The radiation from the synthesizer is then coupled to a Virginia Diodes (VDI) solid-state active multiplier chain, providing continuous coverage across the 80–1600 GHz frequency range. The radiation is directed through a Pyrex tube of 3 m in length and 5 cm in diameter and is detected by a cryogenic-free magnetically enhanced InSb hot-electron bolometer (QMC Instruments Ltd.).

The frequency modulated signal is obtained by encoding the synthesizer radiation with a modulated sine-wave having a modulation width ranging from 100 to 600 kHz.
A lock-in amplifier (SR830, Stanford Research Systems) is used for demodulating the detector output at twice the modulation frequency ($2f$) resulting in a second derivative profile of the absorption signal recorded by the computer-controlled acquisition system. $Z$-apn was measured letting its vapours directly flow through the absorption cell which was continuously pumped and kept at about 3-4 mTorr (0.5\,Pa) and 300K.
\\
%%%%%%%%%%%%%%%%%%%%%%%%%%%%%%%%%%%%%%%%%%%%%%%%%%%%%%%%%%%%%%%%%%%%%%%%%%%%%%%%%%%%%%%%%%%%%%%%%%%%%%%%%%%%%%%%%%%%%%%%%%%%%%%%%%%%%%%%%%%%%%%%%%%%%%
\section{Molecular properties}
\begin{figure}[tb]
 \centering
 \includegraphics[width=0.51\textwidth]{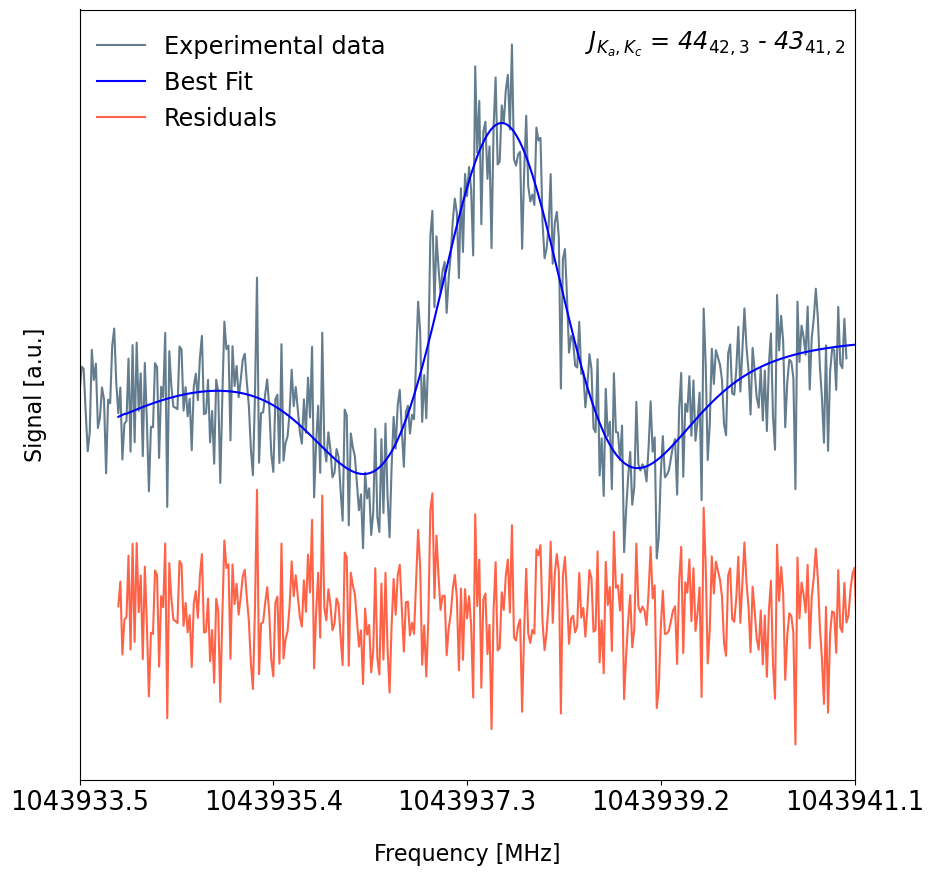}
 \caption{A detail of a $b$-type \textit{Z}-apn transition recorded at 1.04 THz. The line profile is a Voigt profile. In grey is the experimental data, in blue is the best fit, and in red are the residuals. Integration time: 4 minutes; modulation width: 800 kHz; frequency step: 30 kHz. }
    \label{fig:THz}
\end{figure} 

3-Amino-2-propenenitrile (\ch{H2N-CH=CH-C+N}) is produced by the adduction of two ISM abundant molecules, ammonia (NH$_3$) and cyanoacetylene (\ch{HC+C-C+N}). The reaction occurs both in solution \citep{Xiang_1994} or in gas phase \citep{Guillemin_1998}, and yields in a yellow-brown mixture of \textit{E} and \textit{Z} isomers in a 1:1 ratio. After vacuum distillation at 63°C and 0.1 mbar, a colorless mixture of \textit{E} and \textit{Z} isomers was obtained in a 5:95 ratio \citep{Benidar_2005}.
Apn is a disubstituted ethene (\ch{H2C=CH2}), whose substituents are an amino (\ch{-NH2}) and a cyano group (\ch{-CN}). Depending on the orientation of these two groups with respect to the \textit{sp}$^2$ hybridized carbons, apn leads to two isomers. The most stable is denominated as $Z$-, and has the two sub-units oriented on the same side; $E$-apn, has the \ch{-NH2} and \ch{-CN} moieties opposed to each other and is the less energetically favoured \citep{Askeland_2006}. Figure~\ref{fig:isom_energy} shows the energy diagram of the two isomers.

\begin{figure*}[tb]
 \centering
 \includegraphics[width=1\textwidth]{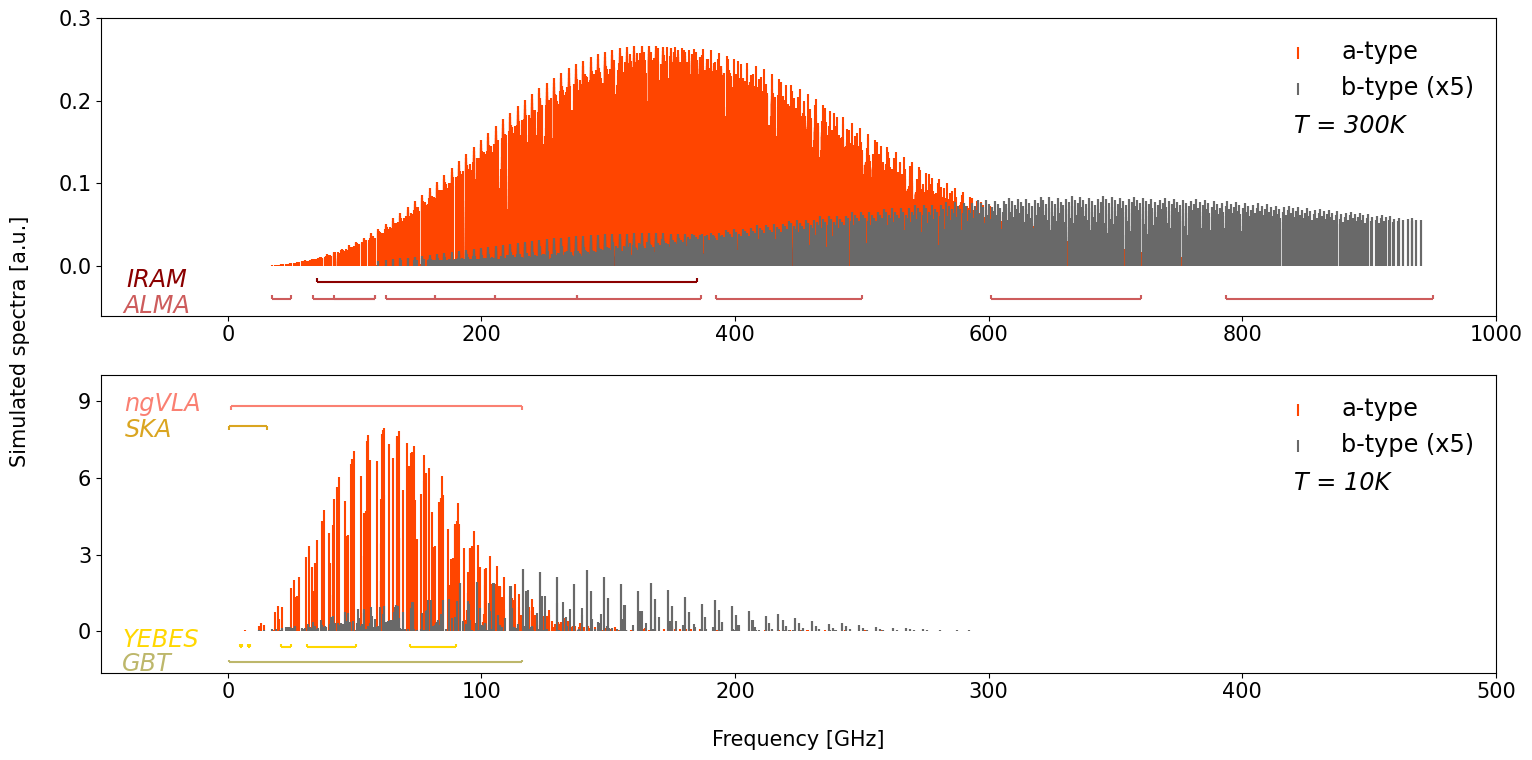}
 \caption{Simulated spectra of $Z$-apn. In red and gray, a-type and b-type transition spectra, respectively. The upper and lower panels are showing the spectra computed at 300\,K and 10\,K, respectively. The x- and y-axis values were chosen to take into account the shift in the Planck function. To note, the spectrum at 300 \,K has a frequency span of three times and an intensity of one-thirtieth compared to the spectrum at 10 \,K. [a.u.] stays for arbitrary units.}
    \label{fig:cat}
\end{figure*} 

In order to estimate this energy difference, we herein conducted an all-electron ab-initio calculations at couple-cluster (CC) level of theory comprising a single, double, and a perturbed triple excitation treatment, i.e. CCSD(T). Employing a cc-pVTZ basis set, the estimated energy difference $\Delta E$ between $E$ and \textit{Z}-apn is $\Delta E=6.63$ kJ mol \textsuperscript{-1} (798K). These same ab-initio calculations allowed us to predict dipole moment components of $E$-apn, extending the contribution of \citet{Askeland_2006} to \textit{Z}-apn alone. The values of both isomers are reported in Table \ref{tab:dipole}, and are expected to be $\mu_a = 5.92$ D, $\mu_b = 0.52$ D and $\mu_c = 0.82$ D and $\mu_a = 4.93(4)$ D and $\mu_b = 0.86(6)$ D, for $E$- and $Z$-apn, respectively. For all calculations we used CFOUR quantum chemistry package \citep{Matthews_2020}. For the most energetically favoured isomer the inertial defect $\Delta = 0.0117$ is also presented, and indicates an almost molecular planarity, with $\Delta$ given by:
\begin{equation} \label{eq:Delta}
    \Delta = I_c-I_b-I_a
\end{equation}

\noindent 
$Z$- and $E$- isomers are both prolate symmetric top rotors and exhibit a Ray's asymmetry parameter $\kappa \approx -0.82$ and $\kappa \approx -0.99$, respectively \citep{Ray_1932}. In the current work, we will focus only on the $Z$- isomer, since this is the only species detected in our laboratory measurements.
\begin{deluxetable}{llcllllr@{.}lllllr@{.}l}[b]
\tablecolumns{6} 
\tablewidth{0pc}
\tablecaption{Relative energies and dipole moment components of the most stable apn isomers.}
\label{tab:dipole}

\tablehead{Parameter && unit &&&&& \multicolumn{2}{c}{\textit{Z}} &&&&& \multicolumn{2}{c}{\textit{E}}}
\startdata
   $E$           &&kJ/mol  &&&&&  0&0    &&&&&    6&63   \\
   $\mu_a$       &&  D     &&&&&  4&93   &&&&&    5&92   \\
   $\mu_b$       &&  D     &&&&&  0&86   &&&&&    0&52   \\
   $\mu_c$       &&  D     &&&&&  0&0    &&&&&    0&82   \\
   $\kappa^a$    &&        &&&&&  0&82   &&&&&    0&99   \\
\enddata
\begin{itemize}
\item [$^a$]$\kappa$ is the asymmetric rotor parameter, given by $(2B-A-C)/(A-C)$ \citep{Gordy_1984}. $\mu_x$ is the dipole moment with respect to the $x$ inertia axis.
\end{itemize}
\end{deluxetable}

\begin{deluxetable}{llccc}[b]
\tablecolumns{4} 
\tablewidth{0pc}
\tablecaption{Transitions and components recorded for Z-apn in this work.}
 \label{tab:transitions}
\tablehead{Type & Branches$^1$ & No. of components & No. of lines}
\startdata
   $a$-type       &  $^aR_{0,+1}$   &  492   &   390   &\\[0.5ex]
   \\
   $b$-type       &  $^bR_{+1,+1}$  &  46    &    35   &\\[0.5ex]
                  &  $^bR_{+1,-1}$  &  32    &    15   &\\[1.0ex]
                  &  $^bR_{-1,+1}$  &   7    &     7   &\\[0.5ex]
                  &  $^bQ_{+1,-1}$  & 110    &    94   &\\[0.5ex]
\enddata
\begin{itemize}
  \item [$^1$] The branch symbol $^{x}M_{\delta_{K_a},\delta_{K_c}}$ refers to the transition type of an asymmetric rotor. x stays for the component of the dipole moment, $M = P, Q, R$ for transitions with $\Delta J = $−$1, 0, $+$1$, respectively, and $\delta K_a$ and $\delta K_c$ for the (signed) change of the \textit{K$_a$} and \textit{K$_{c}$} pseudo-angular quantum numbers \citep{Gordy_1984}.
 \end{itemize}
\end{deluxetable}

%########################################################################################

\section{Analysis} \label{sec:spec}

The high-resolution rotational spectrum of \textit{Z}-apn was herein recorded from 80 GHz to 1.04 THz. 541 new lines were acquired, for a total of 687 components. 492 and 195 transitions belong to the $a$-type and $b$-type spectrum transitions, respectively. For the former, $^aR_{0,+1}$ branches were recorded, while for the latter both $R$ and $Q$ branches were observed. A summary of the different branches recorded is reported in Table \ref{tab:transitions}.
The principal quantum number $J$ ranged from 12 to 117, and the first pseudo quantum number $K_a$ from 0 up to 42.

Whether nitrogen is present in a molecule, quadrupole nuclear hyperfine splitting emerges due to the coupling between the \textsuperscript{14}N nuclear quadrupole moment $I=1$ and the end-over-end molecular rotation.
The hyperfine splitting of a given transition goes with $\Delta_Q \sim -6K^2/J^3$ \citep{Townes_1975}. The transition with the lower principle quantum number recorded in the present work has $J = 12$, around 80 GHz, and no hyperfine features were present. Hyperfine parameters were hence not improved or affected by our measurements, therefore those determined by \citet{Askeland_2006} were maintained.

Most of the transitions were recorded at a pressure of $\sim4$ mTorr at 300 K, but in a few cases, some lines were blended. Among the various contributions that affect line shape and line width\footnote{An important contribution arises due to pressure broadening, other minor ones for power effects, collisions between molecules and cell walls, and distortion introduced by modulators, detectors, and amplifiers.}, a substantial role is played by the Doppler broadening effect, a consequence of the Maxwellian distribution of molecular velocities. 
For electromagnetic radiation tuned at the molecular resonance $\nu_1$, only those molecules having no velocity component along the line of incident radiation will resonate with such frequency. Instead, molecules having a non-zero velocity component along this same axis (in the case in which $\varv_z\ll c$, where c = speed of light), will resonate with frequencies given by
\begin{equation}\label{eq:Dopp} 
    \nu=\nu_1 \bigg(1+\frac{\varv_z}{c}\bigg)
\end{equation}

\noindent The radiation having frequency $\nu$ will then resonate with all those molecules having a velocity along the radiation pathway of 
\begin{equation}\label{eq:v_z} 
    \varv_z=\bigg(\frac{\nu-\nu_1}{\nu_1}\bigg)c
\end{equation}

\begin{deluxetable*}{lr@{.}lcccr@{.}llr}[tb]
%\tablecolumns{9} 
\tablewidth{30pc} 
\tablecaption{Spectroscopic constants determined for $Z$-apn$^a$.}
\label{tab:par}

\tablehead{\multicolumn{3}{c}{This work} &&&& \multicolumn{4}{c}{Askeland et al. (2006)}  \\
\cline{1-3}\cline{7-10} \\[-1.5ex]
\multicolumn{3}{c}{S-Watson reduction$^b$} &&  unit & \multicolumn{5}{c}{A-Watson reduction$^b$}}
\startdata
  $A$                              &    12583&10077(34)  &&  MHz  &&  12583&056(35)     &&$A$                  \\
  $B$                              &     3766&112892(87) &&  MHz  &&   3766&1252(31)    &&$B$                  \\
  $C$                              &     2896&390356(74) &&  MHz  &&   2896&3784(28)    &&$C$                  
  \\[2ex]                                
  $D_J$                            &        4&234231(24) &&  kHz  &&      4&416(17)     &&$\Delta_J$           \\
  $D_{JK}$                         &      -30&07919(29)  &&  kHz  &&    -31&068(98)     &&$\Delta_{JK}$        \\
  $D_K$                            &       92&21050(82)  &&  kHz  &&     46&0(35)       &&$\Delta_{K}$         \\
  $d_1$                            &       -1&4546068(18)&&  kHz  &&      1&4487(52)    &&$\delta_{J}$         \\
  $d_2$                            &       -0&098562(11) &&  kHz  &&      6&4(14)       &&$\delta_{K}$  
  \\[2ex]                     
  $H_J$                            &        0&0165114(44)&&   Hz  &&  \multicolumn{4}{c}{}                    \\
  $H_{JK}$                         &       -0&111108(72) &&   Hz  &&  \multicolumn{4}{c}{}                    \\
  $H_{KJ}$                         &       -0&41083(56)  &&   Hz  &&  \multicolumn{4}{c}{}                    \\
  $H_K$                            &        2&38799(72)  &&   Hz  &&  \multicolumn{4}{c}{}                    \\
  $h_1$                            &        8&5609(32)   &&  mHz  && \multicolumn{4}{c}{}                     \\
  $h_2$                            &        1&3110(27)   &&  mHz  &&\multicolumn{1}{c}{}&         \cline{1-3}                       \\[-3.5ex]
  $h_3$                            &        0&3110(11)   &&  mHz  &&\multicolumn{1}{c}{}&  \multicolumn{1}{c}{Parameter$^c$}    &&    \multicolumn{1}{c}{Value}\\[2ex]
  $L_J$                            &       -0&05557(22)  &&$\mu$Hz&&\multicolumn{1}{c}{}&          \cline{1-3}                      \\[-3.5ex]
  $L_{JJK}$                        &        0&8048(33)   &&$\mu$Hz&&\multicolumn{1}{c}{}&  \multicolumn{2}{l}{2Bz–Bx–By(MHz)}       &\multicolumn{1}{c}{18503.7}\\
  $L_{JK}$                         &       -6&399(32)    &&$\mu$Hz&&\multicolumn{1}{c}{}&  \multicolumn{2}{l}{Bx–By(MHz)}           &\multicolumn{1}{c}{869.7}  \\
  $L_{KKJ}$                        &        0&03051(27)  &&  mHz  &&\multicolumn{1}{c}{}&  \multicolumn{2}{l}{R5 (kHz)}             &\multicolumn{1}{c}{4.726}  \\
  $L_{K}$                          &       -0&08425(26)  &&  mHz  &&\multicolumn{1}{c}{}&  \multicolumn{2}{l}{R6 (kHz)}             &\multicolumn{1}{c}{5.315}  \\
  $l_1$                            &       -0&03135(16)  &&$\mu$Hz&&\multicolumn{1}{c}{}&  \multicolumn{2}{l}{$s_{111}(A)$}         &\multicolumn{1}{c}{0.024}  \\
  $l_2$                            &        5&24(15)     &&  nHz  &&\multicolumn{1}{c}{}&  \multicolumn{2}{l}{$s_{111}(S)$}         &\multicolumn{1}{c}{0.0005} \\
  $l_3$                            &       -4&681(79)    &&  nHz  &&\multicolumn{1}{c}{}&         \cline{1-3}                       \\[-3.5ex]
  $l_4$                            &       -0&942(21)    &&  nHz  &&  \multicolumn{3}{c}{}                    
  \\[2ex]
  $\chi_{aa}$ (NH$_2$)             &        1&7262       &&  MHz  &&   1&7262       \\
  $\chi_{bb}-\chi_{cc}$ (NH$_2$)   &        6&3919       &&  MHz  &&   6&3919       \\
  $\chi_{aa}$ (CN)                 &       -1&4917       &&  MHz  &&  -1&4917       \\
  $\chi_{bb}-\chi_{cc}$ (CN)       &       -1&6371       &&  MHz  &&  -1&6371       \\
  $\sigma_\text{rms}$              &       \multicolumn{2}{c}{44} &&  kHz           \\
  $\sigma_\text{w}$                &        1&18         &&       &&   2&22         \\
No. of lines                       &\multicolumn{2}{c}{935} &\multicolumn{4}{c}{}&86\\
\enddata
\begin{itemize}
  \item [$^a$] Numbers in parenthesis represent the $1\sigma$ standard deviation of the constant in unit of the last digit. Unitless weighted deviation $\sigma_\text{rms}$ of the fit higher than 1 assigns an uncertainty greater than the 67\% confidence level. \item[$^b$] The parameters employed pertain to the I$^r$ representation. For a direct comparison between S- and A- Watson's reduction, refer to \citet{Gordy_1984}. \item[$^c$] List of parameters used to chose the Hamiltonian reduction type to employ in the fitting procedure. To have a straightforward comparison, nomenclature has been taken verbatim from \citep{Margules_2010}. For our purpose, $B_z$, $B_x$ and $B_y$ correspond to the rotational constants $A$, $B$ and $C$, respectively; $R_5$ and $R_6$ are obtained with CFOUR software and are given by combinations of distortion constants and are used to compute the free $s_{111}$ parameter. The latter take part in the Hamiltonian unitary transformation.\\
\end{itemize}
\end{deluxetable*}

%%%%%%%%%%%%%%%%%%%%%%%%%%%%%%%%%%%%%%%%%%%%%%%%%%%%%%%%%%%%%%%%%%%%%%%%%%%%%%%%%%%%%%%%%%%%%%%%%%%%%%%%%

\begin{figure*}[tb]
 \centering
  \includegraphics[width=5.5cm]{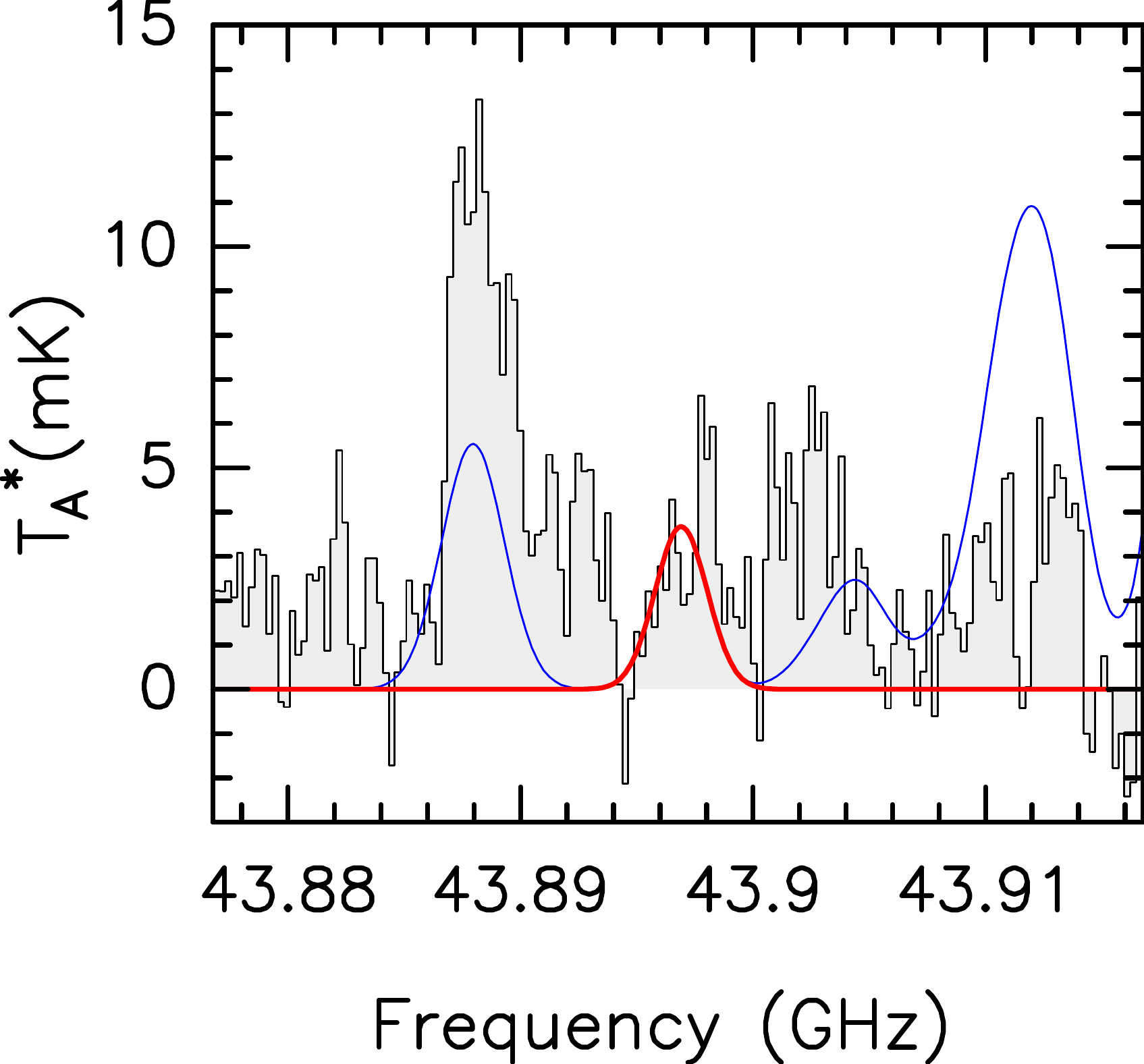}
  \hspace{0.5cm}
 \includegraphics[width=5cm]{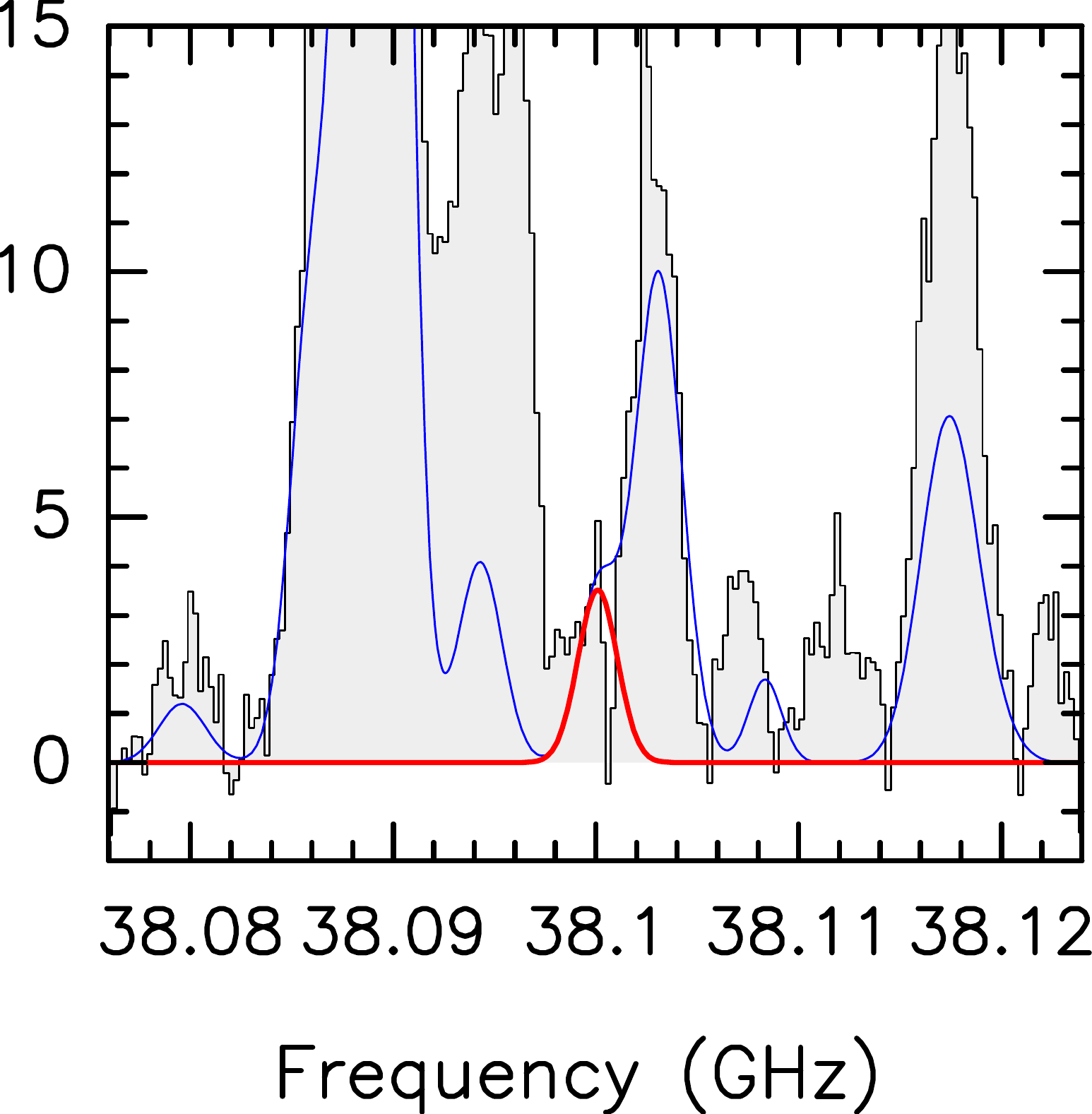}
   \hspace{0.5cm}
  \includegraphics[width=5.2cm]{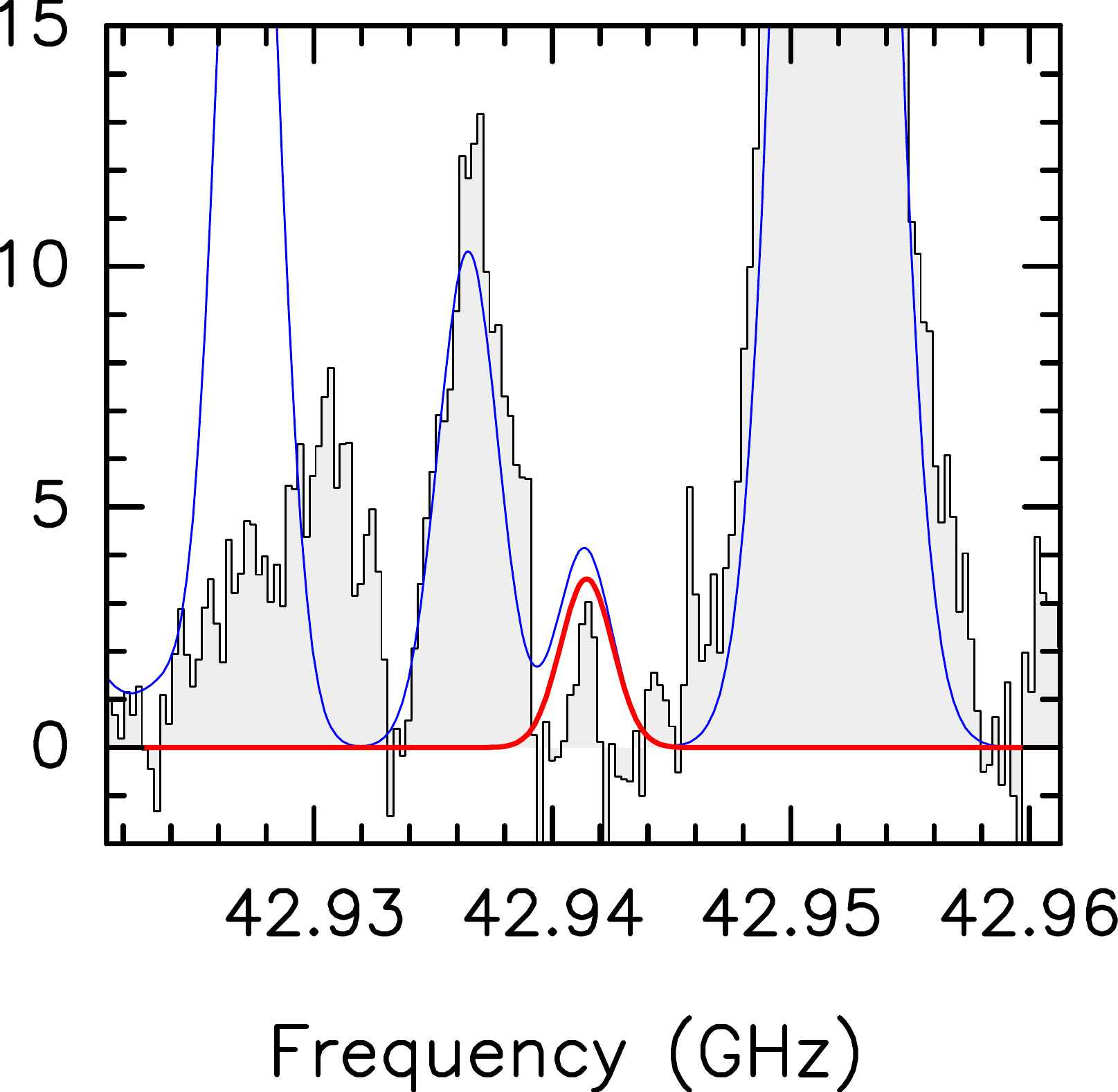}
 \caption{Transitions of $Z$-apn used to compute the column density upper limit towards G+0.693 (see text). The black line and gray histogram show the observed spectrum, the red curve is the LTE synthetic model of $Z$-apn using the column density upper limit derived toward G+0.693, and the blue curve indicates the emission of all the molecules previously identified towards the source.}
    \label{fig:g0693}
\end{figure*} 
%%%%%%%%%%%%%%%%%%%%%%%%%%%%%%%%%%%%%%%%%%%%%%%%%%%%%%%%%%%%%%%%%%%%%%%%%%%%%%%%%%%%%%%%%%%%%%%%%%%%%%%%%

\begin{deluxetable}{cccccc}[tb]
\tablecolumns{6} 
\tablewidth{0pc}
  \tablecaption{Rotational, hyperfine, and vibrational partition functions for \textit{Z-}apn.
  \label{tab:Qr}}
  \tablehead{& \mcl{4}{c}{$Z$-apn}}
  \tablehead
     {$T$[K]&& $Q_\text{rot}$& $Q_\text{HFS}$ && $Q_\text{vib}$}
  \startdata
      2.725  &&     65.85   &    592.64  &&  1.0000 \\
      5.0    &&    162.39   &   1461.56  &&  1.0000 \\
      9.375  &&    415.16   &   3736.44  &&  1.0000 \\
      18.75  &&   1171.51   &  10543.61  &&  1.0000 \\
      37.5   &&   3310.39   &  29793.53  &&  1.0059 \\
      75.0   &&   9362.80   &  84265.20  &&  1.0869 \\
     150     &&  26502.93   &  238526.14 &&  1.5175 \\
     225     &&  48674.53   &  438064.39 &&  2.4062 \\
     300     &&  74545.76   &  670880.28 &&  4.1258 \\
     500     &&  152098.4   &  1368680.8 &&  21.055 \\
    1000     &&  328071.4   &  2951911.5 &&  1381.5 \
    \enddata
\end{deluxetable}

\noindent As theoretically pointed out by \citet{Lamb_1964}, a power saturation of a Doppler-broadened line can produce a sharp dip whenever other broadening factors are small in comparison with Doppler broadening and there is a radiation linearly passing in opposite direction of the gas flow \citep{Winton-Gordy_1970}.
In the present work, the difference of one order of magnitude in pressure obtained passing from the routinely Doppler-limited conditions of few mTorr to 0.2-0.3 mTorr (sub-Doppler regime), allowed to reduce pressure-broadening effects, enabling the saturation of the transition that occurs at frequency $\nu_1$. The subsequent interaction with the radiation reflected by the bottom lens of the absorption cell results in a stimulated emission identifiable by the appearance of a dip in the absorption $2f$ profile, in correspondence of  $\nu_1$. Such low pressures are also important to induce an appreciable separation of the dips from the Doppler profile as well as to minimise possible pressure line shifts \citep{Cazzoli_2003}. 

In Fig. \ref{fig:lamb_dip}, the experimental data recorded under the sub-Doppler regime, are shown in black. In the course of the analysis, the shape of the rightmost line led us to suspect that it was the result of multiple unresolved components, a fact also supported by the predictions obtained from the fit up to that point. By switching to a Lamb-dip regime (black line), it was possible to resolve and assign both lines $J_{Ka,Kc}$ = $75_{3,73}$-$74_{3,72}$ and $76_{2,75}$-$75_{2,74}$, which were then included into the fit. The  measured line widths at full width at half maximum (FWHM) of the transition $75_{3-73}$-$74_{3-72}$ passed from 640 kHz to 97 kHz.  This same procedure was performed to deal with comparable situations, and in combination with sub-Doppler regime analysis allows us to record transitions up to 1 THz. Fig. \ref{fig:THz} shows the transition obtained at the highest frequency, 1.04 THz. Residuals of this line fitting are in good agreement with the Voigt profile fitted using the line profile with our in-house analysis software, that implements the $2f$ Voigt profiles presented by \citet{Dore_2003}.

The new lines were fitted including the ones recorded by \citet{Askeland_2006}, using the SPFIT/SPCAT software. Figures \ref{fig:cat} shows the simulated spectra computed at 300\,K and 10\,K, obtained with the new rotational and distortion constants, presented in Table \ref{tab:par}. The files employed to compute the Z-apn catalogue are provided as Supplementary materials \footnote{The Supplementary materials comprise of the *.lin, *.int, *.fit, *.par, *.var, and *.cat files. These are all available in \textit{Zenodo} \dataset[DOI: 10.5281/zenodo.7904967]{https://doi.org/10.5281/zenodo.7904967}.}. The Z-apn catalogue will be available at CDMS \citep{muller_cologne_2005, endres_cologne_2016} with the name Z-apn.cat. The file format matches the one of CDMS\footnote{https://cdms.astro.uni-koeln.de} and JPL\footnote{https://spec.jpl.nasa.gov/ftp/pub/catalog/catform.html} \citep{pickett_1998}.  In contrast to the latter work, we used the S Watson's reduction asymmetric Hamiltonian, maintaining the $I^r$ representation. This was done to keep the free parameter $s_{111}$, which takes part in the unitary transformation of the Hamiltonian, the lowest as possible, as suggested by \citet{watson_1977b}. This way, the convergence of the fit is sped up and additional terms in the Hamiltonian can be avoided \citep{Margules_2010}. In the sub-table of Table \ref{tab:par} the values used to determine the constants $s_{111}(A)$ and $s_{111}(S)$ are listed, showing a difference of 50 between the former and the latter, respectively. 
The fit was performed by assigning transitions with gradually increasing quantum numbers $J$ and $K_a$. Distortion constants were added in the process in order to account for the divergence due to their increasing values and the corresponding frequencies.

All 687 rotational transitions, including the ones observed in the previous microwave experiment, were reproduced by our model with a final rms uncertainty of 44 kHz. The uncertainty of the new rotational constants has been decreased by a factor of 1000 for $A$, and more than 40 for both $B$ and $C$.
A full set of quartic, sextic and octic distortion constants\footnote{Due to fit, the only excluded octic constants was $l_2$.} with an average uncertainty of $0.003$\%, $0.1$\% and 1\%, respectively, were determined. For the old lines, the same experimental uncertainties attributed by the authors $\sigma$= 3 kHz and $\sigma$=50 kHz were used, for the hyperfine transitions recorded with the Fourier transform spectrometer and Stark spectrometer, respectively. Depending on experimental conditions, line-width and S/N ratio, $\sigma$ ranging from 25 to 50 kHz, were assigned for the transitions recorded in this work.

%and its catalog will be available at CDMS\footnote{https://cdms.astro.uni-koeln.de} \citealt{Muller_2005, Endres_2016} with the name {\fontfamily{cmtt}\selectfont Z-APN.cat}. The file format matchs the one of CDMS and JPL\footnote{https://spec.jpl.nasa.gov/ftp/pub/catalog/catform.html} \citep{pickett_1998}.
Rotational ($Q_{rot}$) partition functions for $Z$-apn are provided in Table \ref{tab:Qr}. The values have been computed for $T=2.725-1000$\,K and are obtained by direct summation over the rotational or hyperfine levels, whose energies have been accurately determined from the spectral analysis. The vibrational partition functions $Q_\text{vib}$ have been computed using harmonic frequencies at the  CCSD(T)/cc-pVTZ level of theory, calculated in this work using CFOUR software \citep{Matthews_2020}. Ab-initio calculation to assist the measurements to determine also the rotational constants of \textit{E}-apn are planned.

\indent\indent
%%%%%%%%%%%%%%%%%%%%%%%%%%%%%%%%%%%%%%%%%%%%%%%%%%%%%%%%%%%%%%%%%%%%%%%%%%%%%%%%%%%%%%%%%%%%%%%%%%%%%%%%%

\section{Interstellar search towards the G+0.693-0.027 molecular cloud}\label{sec:obs}
	
%\textcolor{red}{Victor's text goes here}
%\textcolor{red}{V: working on it}	

We searched for $Z$-apn towards the G+0.693-0.027 molecular cloud, located in the Sgr B2 region of the Galactic Center. Several molecular species, including cyanides and amines, have been recently detected for the first time toward this cloud (see e.g. \citealt{rivilla2019abundant,rivilla2020b,bizzocchi2020,rivilla2021a,rivilla2021b,rivilla2022a,rivilla2022b,rodriguez-almeida2021a,rodriguez-almeida2021b,zeng2021,jimenez-serra2022}).
We used a sensitive unbiased spectral survey performed with the Yebes 40m (Guadalajara, Spain), and the IRAM 30m (Granada, Spain) telescopes. The position switching observations were centered at $\alpha$(J2000.0)=$\,$17$^{\rm h}$47$^{\rm m}$22$^{\rm s}$, $\delta$(J2000.0)=$\,-$28$^{\circ}$21$^{\prime}$27$^{\prime\prime}$. 
%The Yebes 40m observations covered a spectral range from 31.075 GHz to 50.424 GHz, while
%The noise of the spectra depends on the frequency range, with values in antenna temperature ($T_{A}^{*}$) as low as 1.0 mK, while in some intervals it increases up to 4.0$-$5.0 mK.
%the IRAM 30m observations cover the spectral range 71.770 to 116.720 GHz.
%The noise of the spectra (in $T_{A}^{*}$) is 1.3 to 2.8 mK (71$-$90 GHz), 1.5 to 5.8 mK (90$-$115 GHz), and $\sim$10 mK at 115$-$16 GHz. 
The line intensity of the spectra was measured in units of $T_{\mathrm{A}}^{\ast}$ as the molecular emission toward G+0.693 is extended over the beam (\citealt{requena-torres_organic_2006,requena-torres_largest_2008,zeng2018,zeng2020}).
For more detailed information of the observational survey we refer to \citet{rodriguez-almeida2021a} and \citet{rivilla2022c}.

We implemented the spectroscopic entry of $Z$-apn (without considering HFS) from this work into the MADCUBA package{\footnote{Madrid Data Cube Analysis on ImageJ is a software developed at the Center of Astrobiology (CAB) in Madrid; http://cab.inta-csic.es/madcuba/}} (version 28/10/2022; \citealt{martin2019}).
Using the SLIM (Spectral Line Identification and Modeling) tool of MADCUBA, we generated a synthetic spectra of $Z$-apn under the assumption of Local Thermodynamic Equilibrium (LTE), and compare with the observed spectra. We used typical values of the physical parameters found in G+0.693:
$T_{\rm ex}$=8 K, v$_{\rm LSR}$=69 km s\textsuperscript{-1} and FWHM=18 km s\textsuperscript{-1}. 
We show in Figure \ref{fig:g0693} the brightest transitions according to the LTE model that fall in the spectral range covered by the survey\footnote{We have excluded the transition 8$_{0,8}-$7$_{0,7}$ at 49.621278 GHz, because it is completely blended with a brighter transition of Z-cyanomethanimine (Z-HNCHCN; \citealt{rivilla2019abundant}).}: 7$_{0,7}-$6$_{0,6}$ at 43.896761 GHz, 
6$_{0,6}-$5$_{0,5}$ at 38.09996 GHz, and
7$_{1,7}-$6$_{1,6}$ at 42.94129 GHz. 
Although the LTE model is compatible with the observed spectra, we did not detect clear unblended transitions that would allow a secure detection of the molecule. For this reason, we report here an upper limit for its molecular abundance. To derive it we have used the higher value of the column density ($N$) whose LTE model is compatible with the observed spectrum, shown by red curve in Figure \ref{fig:g0693}). We obtained $N<$ 6.6$\times$10$^{12}$ cm$^{-2}$, which means an abundance compared to H$_2$ of $<$ 4.9$\times$10$^{-11}$, using $N$(H$_{2}$)=1.35$\times$10$^{23}$ cm$^{-2}$ (\citealt{martin_tracing_2008}).
%%%%%%%%%%%%%%%%%%%%%%%%%%%%%%%%%%%%%%%%%%%%%%%%%%%%%%%%%%%%%%%%%%%%%%%%%%%%%%%%%%%%%%%%%%%%%%%%%%%%%%%%%

\section{Discussion}\label{sec:disc}

In this work we focused on the ground vibrational state of the lowest energetic isomer $Z$-apn. 
As it was already seen in the cyanides detected in G+0.693, the isomers ratio is unfavourable for the highest energy isomer (see i.e. \citealt{rivilla2019abundant}). Therefore, the probabilities to detect the highest energy isomer $E$-apn, instead of the most stable $Z$- isomer, are lower. Nevertheless, as explained by \citealt{garciadelaconcepcion2021}, while this is true for cyanomethanimine and propynimine ($Z/E>1$) it does not apply to ethanimine ($Z/E<1$). The additional spectroscopic characterisation of the most energetic isomer might help to shed light on the isomer ratio of this molecule in the ISM and on whether higher sensitivity measurements might be available.
%and hopefully understand whether this discrepancies are due to non-thermochemical processes, i.e. in the case of trans-methyl formate (\ce{HCOOCH3}, \citealt{Neill_2012}) and formic and thioformic acids (HCOOH and HC(O)SH, respectively \citealt{GarciadelaConception_2022}).
%A new set of ab-initio calculation constants will guide this search employing the Free-Unit Jet experiment. The latter, exploiting an adiabatic expansion of the molecular jet, has the effect of lowering the rotational temperature at temperatures ranging from $10$ to $15$ K. Like any black body, the shift of Planck's law with low temperatures induces a narrowing of the function profile width, which results in a reduction in the value of the associated partition function. The effect is an increase in the population of the most energetic rotational levels, i.e. the intensity of the low-frequency lines.

As far as rotational spectroscopy is concerned, any molecular system can be described by its angular momentum. The latter is inversely proportional to the rotational constants $A$, $B$ and $C$, determined for $Z$-apn, and to be determined for $E$-apn. For spectroscopic complex molecules such as iCOMs, the associated high moment of inertia results in low rotational constants. In the case of $Z$-apn, the fundamental 1-0 transition, given as for all prolate asymmetric rotors by the sum of the rotational constants $B$ and $C$, is at approximately 6 GHz. Due to the similarity in geometry and atom compositions, one can expect a similar value for $E$ isomer. Being the fundamental rotational level, it has no associated centrifugal distortion, and it is the ideal target when beginning to assign a spectrum for which no prior experimental information is available. For $Z$-apn, it was possible to work nimbly with the CASAC at 300 K because we started from \citet{Askeland_2006}'s work, but for $E$-apn, the same does not apply. The analysis would start from rotational constants obtained from ab-initio calculations, which always have an associated intrinsic uncertainty. On top of this, the effects of centrifugal distortions taking place in higher quantum number transitions make the identification very complex.
Therefore, the Free-Unit Jet experiment is the ideal candidate to begin the assignment of complex molecules  starting from ab-initio calculations, for molecules such as $E$-apn.

The upper limit of $N<$ 6.6$\times$10$^{12}$ cm$^{-2}$ herein obtained (abundance compared to H$_2$ of 4.9$\times$10$^{-11}$, with $N$(H$_{2}$)=1.35$\times$10$^{23}$ cm$^{-2}$) enables the comparison between this upper limit of $Z$-apn with the molecular column densities of simpler cyanides and amines already detected towards G+0.693. Cyanides with two carbon atoms, such as vinylcyanide (\ce{C2H3CN}) and ethylcyanide (\ce{C2H5CN}), are more abundant than $Z$-apn by factors of $\gtrsim$14 and $\gtrsim$6, respectively (\citealt{zeng2018}), while the two-carbon amines vinylamine (\ce{C2H3NH2}) and ethylamine (\ce{C2H5NH2}) are more abundant by factors of $\gtrsim$7 and $\gtrsim$4, respectively (\citealt{zeng2021}). This indicates that the increase of molecular complexity that involves an additional carbon and nitrogen atom implies a drop in molecular abundance of at least one order of magnitude, as it has been found for other molecular families such as alcohols (\citealt{jimenez-serra2022}) and thiols (\citealt{rodriguez-almeida2021a}).

	\section{
Conclusion}\label{sec:concl}

We herein characterized the high-resolution rotational spectrum of $Z$-apn from 80 GHz to 1.04 THz. In the experiment we employed both Doppler and sub-Doppler regime measurements, allowing us to record 541 new lines. With the new $Z$-apn catalogue produced making use of the new rotational constants we searched for this molecule in the G+0.693 molecular cloud. Under LTE condition, we report an upper limit for its column density of $N<$ 6.6$\times$10$^{12}$ cm$^{-2}$, which means an upper limit abundance compared to H$_2$ of 4.9$\times$10$^{-11}$. Further analysis might include the characterisation of the $E$-apn and the search for both isomers in colder sources as well, i.e. L1544 or L183, to help us to better understand the underlying astrochemical link existing between different phases of the star formation regions.
	%---------------------------------------------------------------

	\begin{acknowledgements}
    We gratefully acknowledge the Max Planck society for the financial support. We thanks Yebes 40m telescope staff for help during the different observing runs. The 40m radio telescope at Yebes Observatory is operated by the IGN, Ministerio de Transportes, Movilidad y Agenda Urbana. V.M.R. has received support from the project RYC2020-029387-I funded by MCIN/AEI /10.13039/501100011033. J.-C. G. acknowledges support by the Centre National d’Etudes Spatiales (CNES) and by the Programme National Physique et Chimie du Milieu Interstellaire (PCMI) of CNRS/INSU with INC/INP co-funded by CEA and CNES. I.J-.S and J.M.-P. acknowledge funding from grant No. PID2019-105552RB-C41 awarded by the Spanish Ministry of Science and Innovation/State Agency of Research MCIN/AEI/10.13039/501100011033.
    \end{acknowledgements}
	
	%---------------------------------------------------------------
 
\bibliography{APN}
\bibliographystyle{aasjournal}

    %---------------------------------------------------------------

\end{document}